\documentclass[journal, onecolumn]{IEEEtran}
\usepackage{xcolor}
\usepackage{verbatim}
 \usepackage{multirow}

\usepackage{units}

\usepackage{algorithmic}
\usepackage{algorithm}
\usepackage{array}
\usepackage[caption=false,font=normalsize,labelfont=sf,textfont=sf]{subfig}
\usepackage{textcomp}
\usepackage{stfloats}
\usepackage{url}
\usepackage{verbatim}
\usepackage{graphicx}
\usepackage{cite}

\usepackage[utf8]{inputenc} 
\usepackage[T1]{fontenc}
\usepackage{url}
\usepackage{ifthen}
\usepackage{cite}
\usepackage[cmex10]{amsmath} 

\begin{document}
\title{
Computational Complexity-Constrained Spectral Efficiency Analysis for 6G Waveforms
} 
 \author{%
   \IEEEauthorblockN{Saulo Queiroz\IEEEauthorrefmark{1},
                     João P. Vilela\IEEEauthorrefmark{2}\IEEEauthorrefmark{3},
                     Benjamin Koon Kei Ng\IEEEauthorrefmark{4},
                     Chan-Tong Lam\IEEEauthorrefmark{4}, and
                     Edmundo Monteiro\IEEEauthorrefmark{3}
}
   \IEEEauthorblockA{\IEEEauthorrefmark{1}%
Academic Department of Informatics, Federal University of Technology (UTFPR),\\ Ponta Grossa, PR, Brazil.
                    Email: sauloqueiroz@ufpr.edu.br\\}
   \IEEEauthorblockA{\IEEEauthorrefmark{2}%
CRACS/INESCTEC, and Department of Computer Science, Faculty of Sciences,\\ University of Porto, 4169-007 Porto, 
Portugal.
Email: jvilela@fc.up.pt\\}
   \IEEEauthorblockA{\IEEEauthorrefmark{3}%
Centre for Informatics and Systems  of the University of Coimbra (CISUC),\\ Coimbra, Portugal. Email: \{jpvilela, edmundo\}@dei.uc.pt.\\}
   \IEEEauthorblockA{\IEEEauthorrefmark{4}
Faculty of Applied Sciences,  Macao Polytechnic University, Macao.\\
Email: \{bng, ctlam\}@mpu.edu.mo
}

}


\maketitle

\begin{abstract}
In this work, we present a tutorial on how to account for the computational time complexity overhead of signal processing in the spectral efficiency (SE) analysis of wireless waveforms. Our methodology is particularly relevant in scenarios where achieving higher SE entails a penalty in complexity, a common trade-off present in 6G candidate waveforms. We consider that SE derives from the bit rate, which is impacted by time-dependent overheads. Thus, neglecting the computational complexity overhead in the SE analysis grants an unfair advantage to more computationally complex waveforms, as they require larger computational resources to meet a signal processing runtime below the symbol period.
We demonstrate our points with two case studies. In the first, we refer to IEEE 802.11a-compliant baseband processors from the literature to show that their runtime significantly impacts the SE perceived by upper layers. In the second case study, we show that waveforms considered less efficient in terms of SE can outperform their more computationally expensive counterparts if provided with equivalent high-performance computational resources. Based on these cases, 
we believe our tutorial can address the comparative SE analysis of waveforms that operate under different computational resource constraints.
\end{abstract}

\section{Introduction}
Novel wireless physical (PHY) layer waveforms are expected to reach unprecedented levels of bit rate and spectral efficiency (SE) to meet the traffic demands of beyond-5G scenarios. Towards that goal, it is usual for novel solutions to come at the cost of larger computational complexity overheads {\color{black} in comparison to the classic orthogonal frequency division multiplexing (OFDM) waveform}, thereby impacting other key performance indicators such as manufacturing cost, chip area (portability), and power consumption. Since computational complexity and bit rate/SE often fall under distinct fields of expertise, comparing signal processing techniques that enhance one while compromising the other becomes increasingly difficult. This difficulty hinders the ability to determine whether a certain penalty in computational complexity is worthwhile for a gain in bit rate and SE. 

The complexity-SE trade-off is evident, for example, in index modulation (IM), 
a signal processing technique expected to prominently outperform the SE of 5G~\cite{6Gofdmim-2023}. 
IM can be further combined with other 6G candidate physical layer techniques and waveforms 
like orthogonal time frequency space (OTFS)~\cite{dmofdm-otfs-21}, 
metasurface modulation~\cite{metasuf-im-23}, 
non-orthogonal multiple access (NOMA)~\cite{noma-im-23}, 
faster-than-Nyquist signaling~\cite{7973048}, etc. With such broad flexibility to operate along 
with other signal processing techniques, IM is a good representative of the complexity/SE trade-off 
we are concerned about in this work.

By reviewing the literature, one can identify relatively modest progress towards 
the joint complexity-SE study of waveforms. In~\cite{8352623}, for example,
the authors introduce the `complexity-constrained capacity-achieving NOMA principle' 
to identify design targets in terms of capacity and complexity for a NOMA communication 
system. In~\cite{queiroz-cost-ixs-19},~\cite{queiroz-wcl-19},~\cite{queiroz-access-2020}, 
the authors solve the complexity/SE trade-off for the combinatorial mapper of the original 
OFDM-IM waveform~\cite{basar-ofdmim-globecom-2012}. Through theoretical and experimental tests, 
they demonstrate how to achieve the optimal balance between time complexity and SE for IM waveforms. 
In~\cite{fastenough-2022}, the authors correlate the complexity and throughput of the fast
Fourier transform (FFT) algorithm 
in the context of OFDM waveforms. They show that the FFT complexity nullifies the throughput of OFDM as 
the number of subcarriers grows. These aforementioned references concern the interplay between complexity 
and throughput motivated by trade-offs in specific waveforms and signal processing algorithms.

A remarkable step towards a generic model of capacity and complexity is due 
to~\cite{queiroz2024complim}, in which the authors build 
on~\cite{queiroz-cost-ixs-19},~\cite{queiroz-wcl-19},~\cite{queiroz-access-2020}, 
and \cite{fastenough-2022} to formalize the spectro-computational (SC) analysis. 
The SC analysis is a theoretical framework that enhances classic performance indicators 
of information theory -- such as capacity, bit rate, and SE -- to account for the 
computational complexity overhead of signal processing. In SC terminology, the enhanced 
version of each classic performance indicator is designated by the prefix `SC', e.g., SC 
capacity. These new definitions, however, were mostly intended to enable the formalization 
of a novel capacity regime referred to as comp-limited signals, in which the growth of 
capacity is conditioned by the available amount of computational (rather than spectrum or
 power) resources.

In this work, we rely on the SC framework to present a tutorial about the 
computational complexity-constrained SE analysis of signal waveforms. Such 
joint analysis plays a key role in fairer comparative analyses of waveforms 
constrained by different amounts of computational resources. In these cases, 
neglecting the impact of complexity can lead to unfair comparisons, because 
the superior SE performance of faster signals is conditioned by the allocation 
of more computational resources. If constrained by the same amount of computational
 resources as a slower signal, the fast signal can experience longer signal 
processing delays, thereby hindering its claimed SE if the complexity overhead is 
not neglected.

We illustrate our point through different examples. In one case study, we demonstrate how to compare the SC efficiency of waveforms that require different amounts of computational resources due to different time complexity constraints. We show that waveforms considered less efficient in terms of SE can outperform their more computationally expensive counterparts if provided with equivalent high-performance computational resources. In another case study, we refer to the performance of some OFDM-baseband processors to illustrate how the SE perceived by upper layers can be significantly impacted if the time complexity of the physical layer is not considered. This can occur even when the runtime overhead meets the signal processing requirements of IEEE 802.11a for real-time communications.

The remainder of this work is organized as follows.
In Section~\ref{sec:model}, we present the background of this work.
In Section~\ref{sec:tutorial}, {\color{black} we provide a tutorial on reflecting 
signal processing computational complexity in bit rate and SE analyses. 
Additionally, we explain how to assess these performance indicators under equitable computational 
resource constraints.
In Section~\ref{sec:scanalysis}, we present case studies 
of our tutorial for the OFDM, OFDM-IM, and dual-mode OFDM-IM (DM-OFDM) waveforms and,
in Section~\ref{sec:results}, we present comparative analyses of them based on numerical 
results.
In Section~\ref{sec:conclusion}, we present the conclusion of this work.
For the reader's convenience, Table~\ref{tb:acronyms} lists the acronyms 
and abbreviations used throughout this work, while Table~\ref{tb:symbols} describes
the adopted symbols and notation.
}

\begin{table}[ht]
\caption{\label{tb:acronyms} List of acronyms and abbreviations}
\centering
{\color{black}
\begin{tabular}{|l l|}
\hline
b/s & bits per second\\
b/Hz/s & bits per Hertz per second\\
CP & Cyclic Prefix \\
DM & Dual Mode \\
DM-OFDM & Dual Mode OFDM with IM \\
FFT & Fast Fourier Transform \\
Hz & Hertz \\
IFFT & Inverse FFT \\
IM & Index Modulation \\
inst. & computational instructions\\
inst/s & computational instructions per second\\
IxS & Index Selector \\
LLR & Log-Likelihood Ratio \\
MTU & Maximum Transmission Unit \\
NOMA & Non-Orthogonal Multiple Access \\
OFDM & Orthogonal Frequency Division Multiplexing \\
OFDM-IM & OFDM with IM \\
OTFS & Orthogonal Time Frequency Space \\
PHY & Physical \\
Rx & Receiver \\
SC & Spectro-Computational \\
SE & Spectral Efficiency \\
SNR & Signal-to-Noise Ratio \\
THz & TeraHertz \\
Tx & Transmitter \\
\hline
\end{tabular}
}
\end{table}

\begin{table*}[ht!]
\caption{\label{tb:symbols}{\color{black} Symbols and Notation.}}
\centering
{\color{black}
\begin{tabular}{|l l|}
\hline
$B$ & Bits per symbol\\
$C(n,k)$ & Binomial coefficient ${n\choose k}=n!/(n-k)!$\\
$g$ & Number of subblocks (IM waveforms)\\
$n$ & Number of subcarriers per subblock (IM waveforms)\\
$k$ & Number of active subcarriers (IM waveforms)\\
$M$ & Constellation order\\
$M_A$ & Constellation order of DM-OFDM (mode A)\\
$M_B$ & Constellation order of DM-OFDM (mode B)\\
$N$ & Number of subcarriers\\
$N_{cp}$ & Length of cyclic prefix\\
$R$ & Bit rate\\
$T_{\text{comp-tx}}$ & Signal processing runtime overhead of PHY transmitter\\
$T_{\text{comp-txofdm}}$ & Signal processing runtime overhead of OFDM transmitter\\
$T_{\text{comp-txdm}}$ & Signal processing runtime overhead of DM-OFDM transmitter\\
$T_{\text{comp-txim}}$ & Signal processing runtime overhead of OFDM-IM transmitter\\
$T_{\text{comp-rx}}$ & Signal processing runtime overhead of PHY receiver\\
$T_{\text{comp-rxofdm}}$ & Signal processing runtime overhead of OFDM receiver\\
$T_{\text{comp-rxdm}}$ & Signal processing runtime overhead of DM-OFDM receiver\\
$T_{\text{comp-rxim}}$ & Signal processing runtime overhead of OFDM-IM receiver\\
$T_{\text{rx}}(N)$ &  Complexity function of a $N$-subcarrier PHY receiver\\
$T_{\text{rxdm}}(N)$ & Complexity function of $N$-subcarrier DM-OFDM receiver\\
$T_{\text{rxdm2}}(N)$ & Complexity function of optimized DM-OFDM receiver~\cite{queiroz-access-2020},~\cite{queiroz-wcl-19}\\
$T_{\text{rxim}}(N)$ & Complexity function of $N$-subcarrier OFDM-IM receiver\\
$T_{\text{rxim2}}(N)$ & Complexity function of optimized OFDM-IM receiver~\cite{queiroz-access-2020},~\cite{queiroz-wcl-19}\\
$T_{\text{rxofdm}}(N)$ & Complexity function of $N$-subcarrier OFDM receiver\\
$T_{\text{sym}}$ & Symbol duration\\
$T_{\text{tx}}(N)$ & Complexity function of a $N$-subcarrier PHY transmitter\\
$T_{\text{txdm}}(N)$ & Complexity function of $N$-subcarrier DM-OFDM transmitter\\
$T_{\text{txdm2}}(N)$ & Complexity function of optimized DM-OFDM transmitter~\cite{queiroz-access-2020},~\cite{queiroz-wcl-19}\\
$T_{\text{txim}}(N)$ & Complexity function of $N$-subcarrier OFDM-IM transmitter\\
$T_{\text{txim2}}(N)$ & Complexity function of optimized OFDM-IM transmitter~\cite{queiroz-access-2020},~\cite{queiroz-wcl-19}\\
$T_{\text{txofdm}}(N)$ & Complexity function of $N$-subcarrier OFDM transmitter\\
$\Delta f$ & Subcarrier spacing\\
$W$ & Bandwidth\\
$\mathcal{I}_{\text{rx}}$ &  Required signal processing power of a PHY receiver (inst/s)\\
$\mathcal{I}_{\text{rx-dm}}$ & Required signal processing power of DM-OFDM receiver (inst/s)\\
$\mathcal{I}_{\text{rx-im}}$ & Required signal processing power of OFDM-IM receiver (inst/s)\\
$\mathcal{I}_{\text{rx-ofdm}}$ & Required signal processing power of OFDM receiver (inst/s)\\
$\mathcal{I}_{\text{tx}}$ &  Required signal processing power of a PHY transmitter (inst/s)\\
$\mathcal{I}_{\text{tx-dm}}$ & Required signal processing power of DM-OFDM transmitter (inst/s)\\
$\mathcal{I}_{\text{tx-im}}$ & Required signal processing power of OFDM-IM transmitter (inst/s)\\
$\mathcal{I}_{\text{tx-ofdm}}$ & Required signal processing power of OFDM transmitter (inst/s)\\
$\mathsf{SC}_{\mathsf{SE}}$ & SE accounting for time complexity\\
$\mathsf{SC}_{\text{R}}$ & Bit rate accounting for time complexity\\
$\mathsf{SC}_{\text{R-dm}}$ & DM-OFDM bit rate accounting for time complexity\\
$\mathsf{SC}_{\text{R-im}}$ & OFDM-IM bit rate accounting for time complexity\\
$\mathsf{SC}_{\text{R-ofdm}}$ & OFDM bit rate accounting for time complexity\\
$\mathsf{SC}_{\text{SE-dm}}$ &  DM-OFDM SE accounting for time complexity\\
$\mathsf{SC}_{\text{SE-im}}$ & OFDM-IM SE accounting for time complexity\\
$\mathsf{SC}_{\text{SE-ofdm}}$ & OFDM-IM SE accounting for time complexity\\
$\mathsf{SE}$ & Spectral efficiency\\
$\mathsf{SE}_{\text{ofdm}}$ & Classic spectral efficiency of OFDM\\
\hline
\end{tabular}
}
\end{table*}

\section{Background}\label{sec:model}
In this section, we review the IM waveforms considered in this work and the
SC framework.

\subsection{Index Modulated Waveforms}\label{subsec:imwaveforms}
The OFDM-IM waveform~\cite{basar-ofdmim-globecom-2012} divides 
an $N$-subcarrier OFDM signal into $g>0$ subblocks having
$n=N/g$ subcarriers each. In each subblock, only $k>0$ out of $n$ subcarriers are
active, yielding a total of $C(n,k)={n \choose k}=n!/(n-k)!$ different waveform 
patterns. A total of $\lfloor \log_2 C(n,k) \rfloor$ bits are mapped (demapped)
to these patterns following the so-called Index Selector (IxS) algorithm of OFDM-IM,
which adds a complexity of $O(gnk)$ in comparison to the classic OFDM waveform.
Moreover, each active subcarrier can be modulated as usual by an $M$-ary constellation 
diagram ($M$ is a power of two), yielding a total of $k\log_2 M + \lfloor \log_2 C(n,k) \rfloor$ per subblock.
Aiming to increase this number of bits, the Dual Mode OFDM-IM (DM-OFDM) transmitter~\cite{mao-dm_ofdm_im-ieeeaccess-2017} works just like OFDM-IM unless by the fact that
no subcarrier is turned off. Instead,
two different constellation diagrams -- we denote as $A$ and $B$ --
are employed to differentiate between active and
non active subcarriers, respectively. Considering that $A$ and $B$ have
$M_A$ and  $M_B$ points (both powers of two), respectively, the total number of bits in the DM-OFDM signal 
is $k\log_2 M_A + (n-k)\log_2 M_B + \lfloor \log_2 C(n,k) \rfloor$ per subblock.

Under the `ideal OFDM-IM setup' (which results from setting $g=1$,
$k=N/2$, and $M=2$), the OFDM-IM signal modulates $N/2 + \lfloor \log_2 C(N,N/2) \rfloor$
bits across all $N$ subcarriers, reaching its maximum SE gain over OFDM. However, this entails
a IxS time complexity of $O(N^2)$~\cite{queiroz-access-2020},~\cite{queiroz-wcl-19}.
Similarly, {\color{black} assuming $M=M_A=M_B=2$ for the sake of convenience, 
the DM-OFDM symbol conveys a total of $N + \lfloor \log_2 C(N,N/2) \rfloor$ bits}.
As we will discuss later in Section~\ref{sec:scanalysis}, this is the highest  
complexity along the OFDM-IM block diagram, outing the limits of the complexity-SE
trade-off of the IM waveforms. For this reason, we will assume the ideal IM setup throughout 
this work unless otherwise stated. For other details about the IM waveforms, please, 
 refer to the cited references.

\subsection{SC Analysis}\label{subsec:scframework}
In this section, we review the performance metrics of~\cite{queiroz2024complim} 
to support the computational complexity-constrained SE 
analysis of Section~\ref{sec:scanalysis}.

{\color{black}
The SC efficiency $\mathsf{SC}_{\mathsf{SE}}$ of a signal as wide as $W$ Hertz (Hz)
is for the SC bit rate $\mathsf{SC}_{\text{R}}$  (a.k.a, SC bit rate, SC throughput)
just as the classic SE $\mathsf{SE}$ is for the classical bit rate $R$.
In other words, considering a symbol duration of $T_{\text{sym}}$ seconds
of a signal carrying $B$ bits, the classic bit rate {\color{black} (in b/second, b/s)}
 and SE are defined as 
\begin{eqnarray}
R &=& B/T_{\text{sym}} \quad \text{b/s,}\label{eqn:r}\\
\mathsf{SE} &=& R/W \quad\quad \text{b/s/Hz.}
\end{eqnarray}
Similarly, the homologous of $R$ and $\mathsf{SE}$ in the SC framework, namely,
$\mathsf{SC}_{\text{R}}$ and $\mathsf{SC}_{\mathsf{SE}}$, are related as follows
\begin{eqnarray}\label{eq:scse}
\mathsf{SC}_{\mathsf{SE}} &=& \mathsf{SC}_{\text{R}}/W \quad \text{b/s/Hz.}
\end{eqnarray}
}

{\color{black} To define $\mathsf{SC}_{\text{R}}$, let us denote
$T_{\text{comp-tx}}$ and $T_{\text{comp-rx}}$ as the signal
processing runtime at the transmitter and 
receiver basedband processors, respectively. From this, 
the SC bit rate $\mathsf{SC}_{\text{R}}$ is}
\begin{eqnarray}\label{eq:scr}
\mathsf{SC}_{\text{R}} &=& \frac{B}{T_{\text{comp-tx}}+ T_{\text{sym}}+T_{\text{comp-rx}}} \quad \text{b/s.}
\end{eqnarray}

{\color{black} Eq.~\ref{eq:scr} is the base} to study whether a signal operates
under the comp-limited capacity regime. That regime concerns about whether
the number of bits per unit of transmission (i.e., symbol) grows asymptotically 
faster than the lower bound complexity required to process the symbol. It 
derives from the fixed signal-to-noise ratio (SNR) capacity regime of Shannon
which assumes that capacity grows linearly with the bandwidth.
Assuming packets of bounded length, i.e. the maximum transmission unit (MTU) of 
above-PHY layer, the number of symbol in a PHY layer burst rapidly decreases in 
the fixed SNR regime, since more bits can be conveyed per symbol as spectrum 
widens. Such a condition meets the scenario expected for 6G and beyond wireless
networks, in which the exploitation of terahertz (THz) and sub-THz frequency
bands might enable extremely fast signals because of the abundant spectrum resources~\cite{9145564}.

\begin{figure}[t]
\centering
  \includegraphics[scale=0.4]{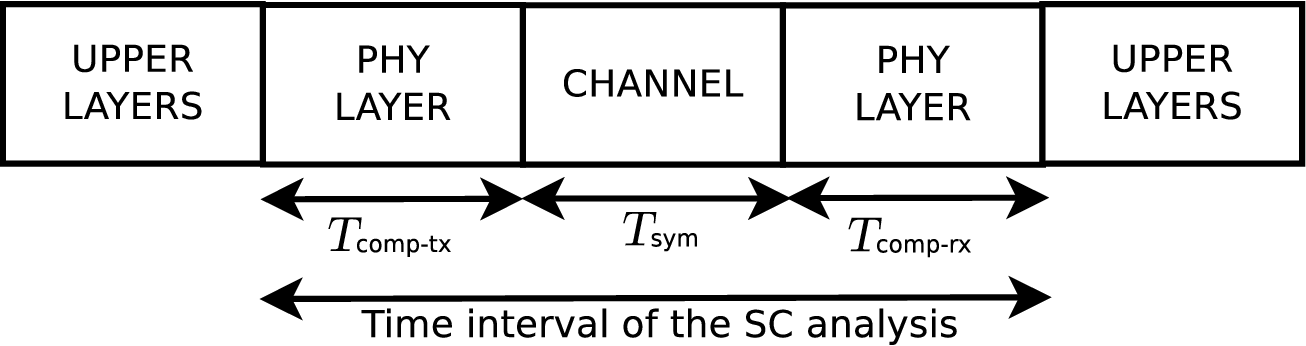}
  \caption{By accounting for the physical layer (PHY) time complexity overheads 
at both transmitter ($T_{\text{comp-tx}}$) and receiver ($T_{\text{comp-rx}}$),
the SC analysis captures the worst-case bit rate of a PHY symbol as perceived by upper layers.}
\label{fig:scheme}
\end{figure}

The study of comp-limited signals considers the maximum delay
a symbol can experience. {\color{black} This matches the delay experienced by the 
first symbol in a burst transmission}.  Such largest delay stems not only from the 
symbol duration time but also from all signal processing delays imposed by the physical layer
before delivering the bits to the above layers. This condition is captured by~\ref{eq:scr}
and is {\color{black}illustrated in Fig.~\ref{fig:scheme}}. Therefore, unless otherwise stated,
the analyses we present throughout this work consider the system premises of the
comp-limited signal regime.

{\color{black}
The signal processing runtime depends on two main factors. The first is the total number 
of computational instructions (inst.) the physical layer must execute on the baseband processor, 
determined by the signal processing algorithms and their input sizes. We denote these 
quantities as \(T_{\text{tx}}(W)\) and \(T_{\text{rx}}(W)\) for the transmitter and receiver, 
respectively. The second factor is the amount of computational resources 
of the baseband processors 
 denoted as \(\mathcal{I}_{\text{tx}}\) and 
\(\mathcal{I}_{\text{rx}}\) for the transmitter and receiver, respectively.
These quantities represent the minimum number of 
instructions per second (inst/s) 
required to process all computational instructions of a signal 
in a runtime no higher than the symbol period.
 Accordingly, the signal 
processing runtime overheads at the transmitter and receiver are defined as follows:
}
\begin{eqnarray}
T_{\text{comp-tx}} &=& \frac{T_{\text{tx}}(W)}{\mathcal{I}_{\text{tx}}} \quad \text{seconds,} \label{eq:tcomptx}\\
T_{\text{comp-rx}} &=& \frac{T_{\text{rx}}(W)}{\mathcal{I}_{\text{rx}}} \quad \text{seconds.} \label{eq:tcomprx}
\end{eqnarray}
Note that the parameters $\mathcal{I}_{\text{tx}}$ and $\mathcal{I}_{\text{rx}}$ capture 
the computational resources (e.g., clock, memory, chip area) allocated to the baseband signal 
processor.


\section{Complexity-Constrained Bit Rate Analysis of Signals}\label{sec:tutorial}
In this section, we {\color{black} present a generic tutorial on how to account
for the signal processing computational complexity overhead when assessing the bit 
rate and SE of a signal. We also show how to ensure fairer comparative analyses of 
signals constrained by different quantities of computational resources.}

{\color{black}
Bit rate and its derived metrics, such as SE, are 
time-sensitive indicators of waveform performance. Comparative analyses of 
different waveforms typically assume the same symbol duration to ensure equitable
 channel time allocation. This assumption directly impacts the computational 
resources assigned to the baseband signal processor for each waveform.

{\color{black} 
A key requirement for baseband processors is to complete the signal processing within the symbol period to maintain real-time communication capabilities~\cite{tan-sora-2011}\cite{6489722},\cite{berger-ofdmintel-2011}. This means that signals with higher computational complexity, requiring more instructions to execute, need larger computational resources to keep the processing runtime within the symbol duration. Neglecting computational complexity and resources in the analysis can introduce bias in comparative rankings based on classic bit rate and SE metrics, favoring more complex waveforms.  The computational resources for these signals would exceed the processing requirements of low-complexity waveforms. These excess resources could allow low-complexity signals to adopt more robust PHY layer settings or advanced signal processing techniques (like stronger error correction codes and more precise channel estimators), ultimately improving bit rate or SE, but at the cost of increased complexity~\cite{queiroz-cost-ixs-19}. Therefore, larger computational resources can enable higher bit rates for low-complexity signals, while high-complexity signals would face longer processing delays (and slower bit rates) if constrained by the hardware limitations of simpler signals.}.

\emph{Our point is that all waveforms should benefit from the same processing 
power capability required by the most computationally complex signal in a 
comparative study}. To achieve this, we adopt the SC framework~\cite{queiroz2024complim}
 and propose a methodology consisted of the following steps 
for comparative performance evaluation of signals constrained by different 
requisites of computational resources:
\begin{enumerate}
\item \textbf{Set the key variables and parameters of the symbol} \\
\begin{enumerate}
 \item \textbf{Symbol duration $T_{\text{sym}}$}: 
As previously discussed, $T_{\text{sym}}$ establishes an
 upper-bound requirement for the signal processing runtime of 
any waveform analyzed in this work. The value of $T_{\text{sym}}$ 
is determined based on channel measurements taken from scenarios that reflect 
the typical environments where the signals are expected to propagate 
(e.g., indoor, outdoor, local area, wide area, etc.). 
Without loss of generality of our tutorial, we adopt a symbol duration of $4$~$\mu$s in 
our case studies, as defined by the IEEE 802.11 standards for indoor wireless 
local area networks.
\item \textbf{Number of bits per symbol $B$, bandwidth $W$, and subcarrier spacing $\Delta f$}: 
As is well known from channel capacity regimes~\cite{queiroz2024complim}, the number of bits that can be reliably transmitted through a symbol primarily depends on the received power and the spectrum bandwidth. For most modern waveforms, $B$ is a function of the number of subcarriers per symbol, $N$, and the constellation order, $M$, which determines the number of bits per subcarrier. For simplicity, one may adopt a single-variable asymptotic analysis. In our case studies, for example, we treat $M$ as a constant parameter, implying that the number of bits per subcarrier is at most $\log_2 M$, and $B$ grows solely with $N$. For OFDM-based waveforms (as considered in the case studies of this work), the symbol bandwidth is given by $W = \Delta f \cdot N$, where $\Delta f$ denotes the subcarrier spacing in Hz. We assume $\Delta f$ to be constant for different values of $N$, which is consistent with practical wireless communication standards (e.g., $\Delta f = 312.5$ kHz in the IEEE 802.11a standard)
\end{enumerate}
\item \textbf{Obtain the complexity function required to process a symbol for each waveform.}\\
Following the assumptions used to define $B$, we also express the computational complexities of all signals as functions of $N$, treating $M$ as a constant.
As is common in algorithm analysis, one may account only for the most computationally intensive algorithm or computational instruction to define the complexity functions $T_{\text{tx}}(N)$ and $T_{\text{rx}}(N)$, representing the computational complexities at the transmitter and receiver, respectively. For instance, the overall complexity of a basic $N$-subcarrier OFDM transmitter (or receiver) can be simplified to the FFT’s complexity (approximately $N \log_2 N$), as it asymptotically dominates the computational cost of the waveform as $N$ grows.
Alternatively, one may consider the complexities of all signal processing algorithms involved in the waveform (examples are provided later). It is important to highlight that any of these assumptions can be modified while preserving the analytical methodology presented in this work.
\item \textbf{Calculate the signal processing power requirement of
the most complex signal.}\\
To identify the most demanding signal in terms of processing
power, compute the quantities $\mathcal{I}_{\text{tx}}$ and
$\mathcal{I}_{\text{rx}}$ for each signal, recalling the constraints
$T_{\text{comp-tx}} \leq T_{\text{sym}}$ and
$T_{\text{comp-rx}} \leq T_{\text{sym}}$.
By allowing each baseband signal processor to operate at the 
maximum permitted processing delay 
(i.e., $T_{\text{comp-tx}} = T_{\text{comp-rx}} = T_{\text{sym}}$) 
and considering Equations~\ref{eq:tcomptx} and \ref{eq:tcomprx} in the context of the most complex 
signal, the required computational resources for the baseband processors
 of the transmitters and receivers of all waveforms in a comparative study are given by:
\begin{eqnarray}
\mathcal{I}_{\text{tx}} &=& \frac{T_{\text{tx}}(N)}{T_{\text{sym}}} \label{eqn:Itx}\quad \text{inst/s,} \\
\mathcal{I}_{\text{rx}} &=& \frac{T_{\text{rx}}(N)}{T_{\text{sym}}} \label{eqn:Irx} \quad \text{inst/s,} 
\end{eqnarray}
respectively.
{\color{black}
\item \textbf{Calculate the signal processing runtime for all waveforms assuming the processing power
required by the most complex signal.}\\
With the quantities $\mathcal{I}_{\text{tx}}$ and $\mathcal{I}_{\text{rx}}$ derived
from the prior step and the complexity functions of each signal, 
compute the signal processing runtime of each waveform according to~\ref{eq:tcomptx} and \ref{eq:tcomprx}.
\item \textbf{Compute the complexity-constrained bit rate and SE of each signal.}\\
Finally, based on the number of bits per symbol of each waveform design and the
 signal processing runtime obtained in the prior step,
compute the complexity-constrained bit rates and SEs of all
waveforms by referring to the~\ref{eq:scr} and \ref{eq:scse},
respectively.
}
\end{enumerate}
Figure~\ref{fig:tutorial} provides a flow diagram summarizing the methodology. 
It highlights the required inputs and the quantities produced at each step of the process. 
Note that this methodology is applicable to any waveform, as long as its number of bits, 
symbol duration, and computational complexity per symbol are defined.}

\begin{figure*}[t]
\centering
  \includegraphics[scale=0.35]{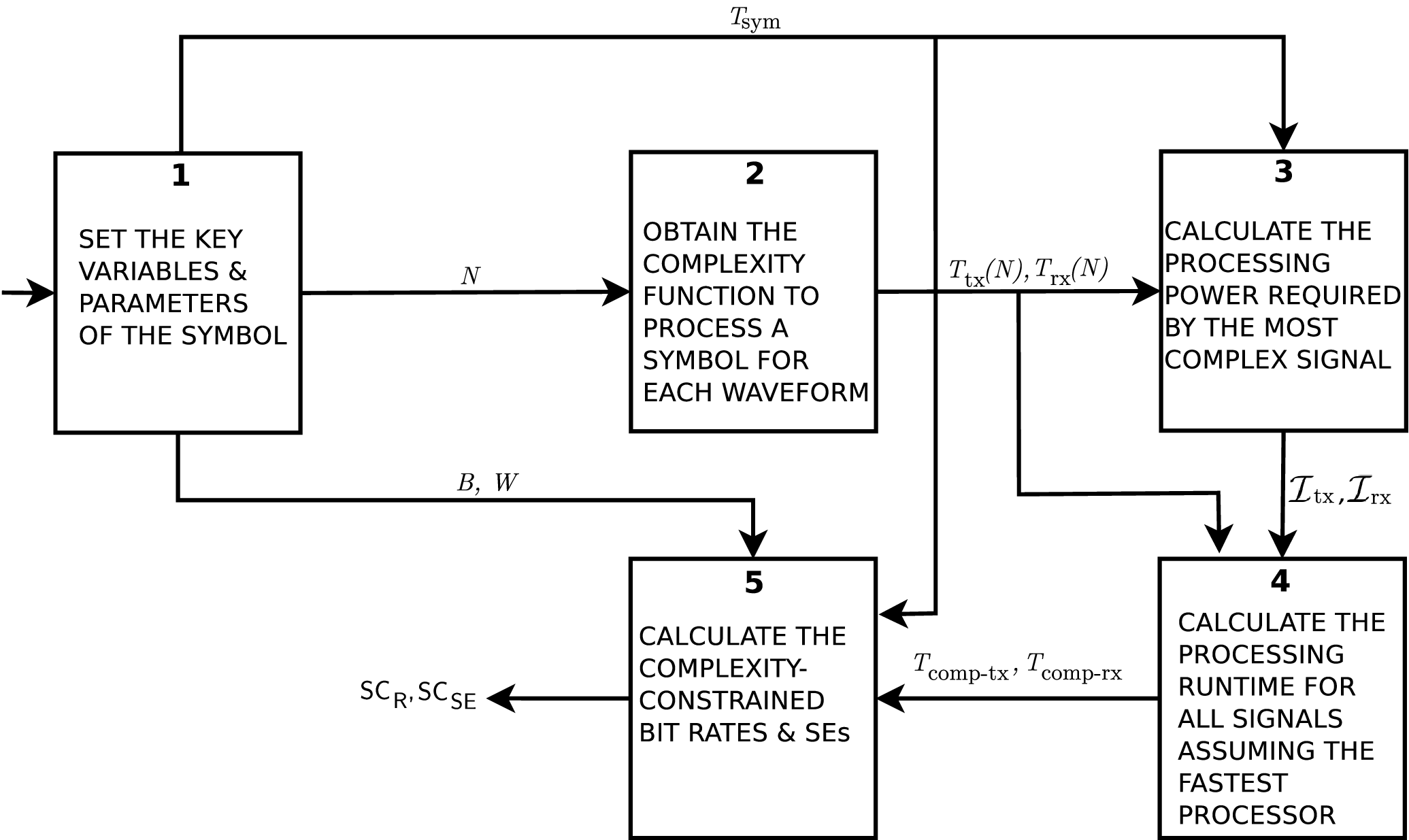}
  \caption{Flow diagram to assess (and compare) the complexity-constrained bit rates $\mathsf{SC}_{\mathsf{R}}$
and SEs $\mathsf{SC}_{\mathsf{SE}}$ of signals constrained by different processing requirements.
Considering an $N$-subcarrier transmitter that modulates $B$ bits in a $W$-Hz signal,
the baseband signal processor  takes $T_{\text{comp-tx}}$ seconds to execute
$T_{\text{tx}}(N)$ computational instructions if empowered by the same processing 
power $\mathcal{I}_{\text{tx}}$ (inst/s)
required by the most complex signal of the comparative study. This runtime overhead
adds to the symbol duration $T_{\text{sym}}$ and the homologous runtime at the receiver  
$T_{\text{comp-rx}}$ to compute $\mathsf{SC}_{\mathsf{R}}$ (b/s) and $\mathsf{SC}_{\mathsf{SE}}$ (b/s/Hz).
}
\label{fig:tutorial}
\end{figure*}

{\color{black}
\section{Case Studies}\label{sec:scanalysis}
In this section,  we exemplify the tutorial of section~\ref{sec:tutorial} 
by introducing case studies for the classic OFDM waveform and the IM variants 
OFDM-IM and DM-OFDM.
}
\subsection{Waveforms Time Complexity}
OFDM and IM waveforms have the same computational complexity except for the
{\color{black} IxS} and the detection algorithms.
As we discussed in Section~\ref{sec:model}, the IxS of the IM waveforms add an extra
complexity of $O(gnk)$ in comparison to OFDM. This overhead verifies at both the
transmitter and the receiver. Thus, considering the ideal IM setup described in 
Section~\ref{sec:model}, the complexity overhead introduced by each IxS is 
$O(1\cdot N\cdot N/2)=O(N^2)$. Recall that the {\color{black} IxS} complexity is zero for OFDM.
Therefore, assuming as negligible both the time complexities of the look-up table 
(de)mapping and the {\color{black} cyclic prefix (CP)} addition/remotion, and omitting 
the implementation-dependent constants,
the overall number of computational instructions performed by the OFDM, OFDM-IM, and 
DM-OFDM transmitters under the ideal IM setup can be given by~\ref{eq:ttxofdm},~\ref{eq:ttxim}, and~\ref{eq:ttxdm}, respectively.
\begin{eqnarray} 
T_{\text{txofdm}}(N) &=&  \overbrace{N\log_2 N}^{\text{IFFT}}  \quad \text{inst.} \label{eq:ttxofdm}\\
T_{\text{txim}}(N) &=& N\log_2 N + \overbrace{N^2}^{\text{IxS}} \quad \text{inst.}\label{eq:ttxim}\\
T_{\text{txdm}}(N) &=& N\log_2 N + N^2 \quad \text{inst.} \label{eq:ttxdm}
\end{eqnarray}
{\color{black}  
Recalling that IM waveforms split $N$ subcarriers into $g$ groups of
$n$ subcarriers each, i.e., $N=gn$, and the ideal setup implies in $M=M_A=M_B=2$, 
the growth of the signal detection complexity solely depends on the number 
of subcarriers rather than on the constellation length $M$. This is due to the
log-likelihood ratio (LLR)-based detection strategy~\cite{basar-ofdmim-globecom-2012}.
In this context, the complexity of detection  for  OFDM, OFDM-IM, and DM-OFDM become
$O(N\log_2 M)=O(N)$, $O(gnM)=O(2N)$\footnote{We deliberately don't suppress from our complexity analyses the constants 
associated to the waveform-related parameters (e.g., $M$, $M_A$, and $M_B$)
because of their impact on the `real-time constraint' of the signal processing.}
and $O(gn(M_A+M_B))=O(4N)$, respectively. }
Thus, the resulting time complexity at their respective receivers are
\begin{eqnarray} 
T_{\text{rxofdm}}(N) &=& \overbrace{N\log_2 N}^{\text{FFT}}+ \overbrace{N}^{\text{LLR}}\quad \text{inst.} \label{eq:trxofdm}\\
T_{\text{rxim}}(N) &=&   N\log_2 N + 2N + \overbrace{N^2}^{\text{de-IxS}}\quad \text{inst.} \label{eq:trxim}\\
T_{\text{rxdm}}(N) &=&   N\log_2N + 4N+N^2 \quad \text{inst.}\label{eq:trxdm}
\end{eqnarray}

It is remarkable to note that the original IxS leads to a computational complexity higher than
the detection and FFT procedures if SE maximizes (i.e., $k=N/2$, $g=1$), although 
some DFT algorithms may outperform the FFT's computational complexity 
for specific scenarios not covered by the case studies of this work~\cite{queiroz-ieeespm-2025}. 
From the general $O(gnk)$ time complexity of the original IxS algorithm, it is easy to conclude 
that \emph{the originally proposed IxS does not run in linear time complexity unless $k=O(1)$}.
In this case, one might give up the optimal SE, since the optimal balance between time complexity 
and SE is achieved if the IM mapper runs in exact $\Theta(N)$ time complexity (though faster algorithms are possible)
and $k=N/2$, as observed in~\cite{queiroz-cost-ixs-19}. Such optimized solution was proposed 
in~\cite{queiroz-access-2020},~\cite{queiroz-wcl-19}, based on which the overall 
computational complexity of the OFDM-IM and DM-OFDM 
transmitters \ref{eq:ttxim}, \ref{eq:ttxdm},
and receivers~\ref{eq:trxim},~\ref{eq:trxdm}, respectively,
improve to
\begin{eqnarray}
T_{\text{txim2}}(N) &=& N\log_2 N + \overbrace{N}^{\substack{\text{Optimized} \\ \text{IxS\cite{queiroz-access-2020},\cite{queiroz-wcl-19}}}} \text{inst.}\label{eq:ttxim2}\\
T_{\text{txdm2}}(N) &=& N\log_2 N + N \quad \text{inst.} \label{eq:ttxdm2}\\
T_{\text{rxim2}}(N) &=&   N\log_2 N + 2N + \overbrace{N}^{\substack{\text{Optimized} \\\text{de-IxS}}} \text{inst.} \label{eq:trxim2}\\
T_{\text{rxdm2}}(N) &=&   N\log_2N + 4N+N \quad \text{inst.}\label{eq:trxdm2}
\end{eqnarray}

Despite of that, we have observed that the IM literature is mostly focused on the
``complexity vs. communication'' trade-off of the detection problem and they usually 
refer to a IxS procedure reminiscent to the non-optimized original IM proposal 
e.g.,~\cite{8681607},~\cite{6Gofdmim-2023}. Due to this, in addition to the
``optimized'' IM, our analyses throughout this work will also consider the time complexity 
of the original IM mapper popularly referred to by the IM literature as given 
in~\ref{eq:ttxim},~\ref{eq:ttxdm},~\ref{eq:trxim}, and~\ref{eq:trxdm}.

\subsection{Baseband Signal Computation Runtime}\label{subsec:tcomp}
The computational resources required by the design of
a given digital baseband processor are constrained by several factors 
like budget, portability, power consumption, runtime constraints, etc. We abstract these
resources by modeling the resulting processing power in algorithmic instructions per second (inst/s).
Thus, let us denote 
$\mathcal{I}_{\text{tx-ofdm}}$, $\mathcal{I}_{\text{tx-im}}$, and $\mathcal{I}_{\text{tx-dm}}$ 
the minimum number of inst/s required by the baseband signal processors of the OFDM, 
OFDM-IM, and DM-OFDM transmitters, respectively. Besides, denote 
$\mathcal{I}_{\text{rx-ofdm}}$, $\mathcal{I}_{\text{rx-im}}$, and $\mathcal{I}_{\text{rx-dm}}$
as the homologous performance indicators for the corresponding receivers, respectively.
{\color{black} Considering the computational complexities of the OFDM, OFDM-IM, and DM-OFDM
transmitters as given by~\ref{eq:ttxofdm}, \ref{eq:ttxim}, and \ref{eq:ttxdm}, respectively, 
the corresponding baseband signal computation time 
of the OFDM, OFDM-IM and DM-OFDM transmitters are given as follows,}
\begin{eqnarray} 
T_{\text{comp-txofdm}}(N) &=& \frac{T_{\text{txofdm}}(N)}{\mathcal{I}_{\text{tx-ofdm}}}  \quad \text{seconds}, \label{eq:tcomptxofdm}\\
T_{\text{comp-txim}}(N) &=& \frac{T_{\text{txim}}(N)}{\mathcal{I}_{\text{tx-im}}}  \quad \text{seconds}, \label{eq:tcomptxim}\\
T_{\text{comp-txdm}}(N) &=& \frac{T_{\text{txdm}}(N)}{\mathcal{I}_{\text{tx-dm}}} \quad \text{seconds}. \label{eq:tcomptxdm}
\end{eqnarray}

Similarly, {\color{black} based on the computational complexities of the
OFDM, OFDM-IM, and DM-OFDM receivers, given by~\ref{eq:trxofdm}, \ref{eq:trxim}, and \ref{eq:trxdm}, respectively,} the baseband signal
computation time at the OFDM, OFDM-IM and DM-OFDM receivers result
\begin{eqnarray} 
T_{\text{comp-rxofdm}}(N) &=& \frac{T_{\text{rxofdm}}(N)}{\mathcal{I}_{\text{rx-ofdm}}}  \quad \text{seconds} \label{eq:tcomprxofdm},\\
T_{\text{comp-rxim}}(N) &=& \frac{T_{\text{rxim}}(N)}{\mathcal{I}_{\text{rx-im}}}  \quad \text{seconds}\label{eq:tcomprxim},\\
T_{\text{comp-rxdm}}(N) &=& \frac{T_{\text{rxdm}}(N)}{\mathcal{I}_{\text{rx-dm}}} \quad \text{seconds}. \label{eq:tcomprxdm}
\end{eqnarray}

In the above equations, we decouple computational complexity (numerator) from the
baseband processing power (denominator). Hence, hardware techniques that speed
up computation (e.g., pipelining, LUTs) can decrease the wall-clock runtime
but cannot decrease time complexity, since this latter quantity solely depends 
on the signal processing algorithms.


\subsection{SC Throughput}
The total number of bits entering the digital baseband processors of the OFDM, OFDM-IM, 
and DM-OFDM waveforms are $N\log_2 M$, $N/2\log_2 M+\lfloor \log_2 {N \choose N/2} \rfloor$,  and
$N/2\log_2 M_A+N/2\log_2 M_B+\lfloor \log_2 {N \choose N/2} \rfloor$, respectively. 
The algorithmic throughput
(also referred to as SC throughput) of OFDM, OFDM-IM and DM-OFDM are respectively
defined as 
\begin{eqnarray}
\mathsf{SC}_{\text{R-ofdm}}(N) &=& \frac{N\log_2 M}{T_{\text{comp-txofdm}}(N)+T_{\text{sym}}+T_{\text{comp-rxofdm}}(N)}\nonumber\\
&&\quad\text{b/s,}\label{eq:scofdm}\\
\mathsf{SC}_{\text{R-im}}(N) &=&    \frac{N/2\log_2 M+\lfloor \log_2 {N \choose N/2} \rfloor}{T_{\text{comp-txim}}(N)+T_{\text{sym}}+T_{\text{comp-rxim}}(N)}\nonumber\\
&&\quad\text{b/s,} \label{eq:scim}\\
\mathsf{SC}_{\text{R-dm}}(N) &=& \frac{N/2(\log_2 M_A+\log_2 M_B)+\lfloor \log_2 {N \choose N/2} \rfloor}{T_{\text{comp-txdm}}(N)+T_{\text{sym}}+T_{\text{comp-rxdm}}(N)} \nonumber\\
&&\quad\text{b/s.} \label{eq:scdm}
\end{eqnarray}

Under the ``ideal setup'' assumed throughout this work, 
one gets $M=M_A=M_B=2$, yielding a total number of bits of
$N$, $N/2+\lfloor \log_2 {N \choose N/2} \rfloor$, and 
$N+\lfloor \log_2 {N \choose N/2}\rfloor$, for OFDM, OFDM-IM, and
DM-OFDM, respectively.

\subsection{Time Complexity-Constrained SE} 
The SE constrained by the time complexity overhead (also referred to as 
SC efficiency, SCE) readily results from the SC throughput. For OFDM, OFDM-IM, and DM-OFDM
it is respectively defined as,
\begin{eqnarray}
\mathsf{SC}_{\mathsf{SE}\text{-}\mathsf{ofdm}}(N) &=&  \frac{\mathsf{SC}_{\text{R-ofdm}}(N)}{W} 
\quad \text{b/s/Hz,} \label{eq:sceofdm}\\
\mathsf{SC}_{\mathsf{SE}\text{-}\mathsf{im}}(N) &=&    \frac{\mathsf{SC}_{\text{R-im}}(N) }{W} \quad \text{b/s/Hz,} \label{eq:sceim}\\
\mathsf{SC}_{\mathsf{SE}\text{-}\mathsf{dm}}(N) &=& \frac{\mathsf{SC}_{\text{R-dm}}(N)}{W} \quad \text{b/s/Hz.}  \label{eq:scedm}
\end{eqnarray}

As discussed in section~\ref{sec:tutorial}, the computation time (of either the transmitter
or the receiver) must not be larger than the symbol duration
to ensure a real-time capable physical layer implementation. Thus, the largest tolerable value for 
the computation time is the symbol duration. Considering this case,
the SCEs of~\ref{eq:sceofdm},~\ref{eq:sceim}, 
and~\ref{eq:scedm} respectively simplify as follow, 
\begin{eqnarray}
\mathsf{SC}_{\mathsf{SE}\text{-}\mathsf{ofdm}}(N) &=&  \frac{\mathsf{SC}_{\text{R-ofdm}}(N)}{3(N+N_{cp})} 
\quad \text{b/s/Hz,} \label{eq:sceofdm2}\\
\mathsf{SC}_{\mathsf{SE}\text{-}\mathsf{im}}(N) &=&    \frac{\mathsf{SC}_{\text{R-im}}(N) }{3(N+N_{cp})} \quad \text{b/s/Hz,} \label{eq:sceim2}\\
\mathsf{SC}_{\mathsf{SE}\text{-}\mathsf{dm}}(N) &=& \frac{\mathsf{SC}_{\text{R-dm}}(N)}{3(N+N_{cp})} \quad \text{b/s/Hz.}  \label{eq:scedm2}
\end{eqnarray}
The aforementioned simplification stems from the fact that
the denominator of the SC throughput rewrites to $3T_{\text{sym}}$ for 
each waveform if the computation time is equal to the symbol duration.
 From this, it results
 $W\cdot 3T_{\text{sym}}=N\Delta f\cdot 3(1/\Delta f + T_{\text{cp}})=3(N+N_{\text{cp}})$.

\section{Numerical Results}\label{sec:results}
In this section, we analyse the OFDM bit rate and SE under the
time computational overheads of specific baseband 
processors (section~\ref{subsec:baseband}). In Section~\ref{subsec:comparison},
we investigate whether the superior SE claimed by IM waveforms do hold
if its time complexity overhead is accounted for the bit rate analysis.
\begin{table}[]
\caption{Complexity-constrained throughput of a IEEE 802.11a WiFi symbol
considering the signal processing overhead of different baseband 
processors ($N=48$, $M=64$).}
\begin{tabular}{|l|cc|c|c|}
\hline
\multicolumn{1}{|c|}{\multirow{2}{*}{Comparison}}                                                  & \multicolumn{2}{c|}{\begin{tabular}[c]{@{}c@{}}Processing\\ delay ($\mu$s)\end{tabular}} & \multirow{2}{*}{\begin{tabular}[c]{@{}c@{}}Symbol\\ duration ($\mu$s)\end{tabular}} & \multirow{2}{*}{\begin{tabular}[c]{@{}c@{}} Throughput \\(b/$\mu$s)\end{tabular}} \\ \cline{2-3}
\multicolumn{1}{|c|}{}                                                                             & \multicolumn{1}{c|}{Transmitter}                   & Receiver                   &                                                                            &                                                                                 \\ \hline
Processor A \cite{6489722}                                                        & \multicolumn{1}{c|}{0.55}                          & 3.59                       & 4                                                                          & 33.8                                                                            \\ \hline
Processor B \cite{berger-ofdmintel-2011}                                          & \multicolumn{1}{c|}{0.55}                          & 3.29                       & 4                                                                          & 36.7                                                                            \\ \hline
\begin{tabular}[c]{@{}l@{}}Ideal processor,\\ Eq.~\ref{eqn:r}\end{tabular} & \multicolumn{1}{c|}{0}                             & 0                          & 4                                                                          & 72                                                                              \\ \hline
\end{tabular}
\label{tb:wifi}
\end{table}

\subsection{Impact of Baseband Runtime on the OFDM Performance} \label{subsec:baseband}
In this subsection, we refer to the OFDM baseband processors of~\cite{6489722} 
and~\cite{berger-ofdmintel-2011} to demonstrate how SE can be severely impacted
if the signal processing time overhead is considered akin the CP time overhead.
For the sake of discussion, let us label these baseband processors as A
and B, respectively. As reported in~\cite{6489722},
processor A achieves a computation time of $T_{\text{comp-txA}}=0.55$~$\mu$s and
$T_{\text{comp-rxA}}=3.95$~$\mu$s at the transmitter and receiver, respectively.
These results are reported for an IEEE 802.11a signal of 54 Mb/s 
(i.e., $M=64$, $N=64$, $T_{\text{cp}}=0.8$~$\mu$s, and $T_{\text{sym}}=4$~$\mu$s).
Under the same bit rate, processor B achieves a runtime of $T_{\text{comp-rxB}}=3.29$~$\mu$s
at the receiver. The runtime $T_{\text{comp-txB}}$ for the transmitter A is not 
reported by~\cite{berger-ofdmintel-2011}.
For this case, in what follows we assume a transmitter empowered by the same baseband 
processor of A, i.e.,
$T_{\text{comp-txB}}=T_{\text{comp-txA}}=0.55$~$\mu$s. These time parameters
and the SC throughput we discuss next are summarized in Table~\ref{tb:wifi}.

In all cases, the real-time runtime signal processing requisite  $T_{\text{comp}}\leq T_{\text{sym}}$
is satisfied by the transmitters and the receivers.
However, the runtime at the receivers A and B are $\approx 4.93\times$ 
and $\approx 4.11\times$ higher than the standard CP overhead of IEEE 802.11a. Besides, the 
respective signal computation runtime of the processors correspond 
to circa of $53\%$ and $49\%$ of the overall time overheads 
introduced by the physical layer, i.e.  $T_{\text{comp-tx}} + T_{\text{sym}} + T_{\text{comp-rx}}$. 
To assess the complexity-constrained bit rate perceived by the upper layers,
we refer to the SC throughput. Recalling that only $48$ (out of the $N=64$) subcarriers convey
useful data bits in IEEE 802.11a and that six bits are sent per subcarrier by 64-QAM points ($M=64$) in
this case,
 one gets a SC bit rate of 
{\color{black}
\begin{eqnarray}
\mathsf{SC}_{\text{R-A}}(48)&=&\frac{48\cdot 6}{0.55+4+3.95}\approx 33.8 \quad \text{b/$\mu$s}, 
\end{eqnarray}
}
for a communication empowered by the baseband processor A. Homologously, the baseband processor B 
achieves a SC throughput 
{\color{black}
\begin{eqnarray}
\mathsf{SC}_{\text{R-B}}(48)&=&\frac{48\cdot 6}{0.55+4+3.29}\approx 36.7 \quad \text{b/$\mu$s}.
\end{eqnarray}
}
In both cases, the OFDM bit rate without the computational overheads 
is 
{\color{black}
\begin{eqnarray}
R(48)&=&\frac{48\cdot 6}{4}=72 \quad \text{b/$\mu$s}, 
\end{eqnarray}
 nearly half of the obtained if
the baseband processor runtime overhead is not neglected.} 
\emph{Therefore, similarly to the CP overhead, the signal processing computational 
complexity overhead can dramatically impair the waveform bit rate even when the
baseband signal processor can run below the symbol period, as required for real-time 
communication performance.}
Consequently, (complexity-constrained) SE impairs accordingly. 

Given the IEEE 802.11a standard bandwidth of $W=20$~MHz, the expected OFDM SE of our case study
impairs from 
{\color{black}
\begin{eqnarray}
 \mathsf{SE}_{\text{ofdm}}&=&\frac{R(48)}{20\text{ MHz}}=\frac{72\text{ b/$\mu$s}}{20 \text{ MHz}}=3.60 \quad \text{b/s/Hz,}
\end{eqnarray}
to 
\begin{eqnarray}
 \mathsf{SC}_{\text{SE-A}}&=&\frac{\mathsf{SC}_{\text{R-A}}(48)}{20\text{ MHz}}=\frac{33.8\text{ b/$\mu$s}}{20 \text{ MHz}} \nonumber\\
&=&1.69 \quad \text{b/s/Hz }, \\
\mathsf{SC}_{\text{SE-B}}&=&\frac{\mathsf{SC}_{\text{R-B}}(48)}{20\text{ MHz}}=\frac{36.7\text{ b/$\mu$s}}{20 \text{ MHz}}\nonumber\\
&\approx& 1.83 \quad \text{b/s/Hz }, 
\end{eqnarray}
for OFDM being processed on baseband processors A and B,} respectively.

\subsection{SE Constrained by Time Complexity: IM vs. OFDM}\label{subsec:comparison}
In this section, we present a comparative time complexity-constrained SE analysis 
among OFDM, OFDM-IM, and DM-OFDM waveforms. 
{\color{black} The analysis follows the steps of Section~\ref{sec:tutorial} assuming the}
`ideal IM setup', i.e., $M=2$ (BPSK modulation), $k=N/2$, and $g=1$ for the IM waveforms. 
{\color{black} We assume the PHY layer parameters of the IEEE 802.11a standard 
for all waveforms, $T_{\text{sym}}=4~\mu$s and $N=64$.}

\subsubsection{Required Computational Resources of the Waveforms}
{\color{black} By referring to~\ref{eqn:Itx} for a symbol duration $T_{\text{sym}}=4$~$\mu$s,
and considering the computational complexities given in~\ref{eq:ttxofdm},
\ref{eq:ttxim}, and \ref{eq:ttxdm} for $N=64$-subcarrier signals, 
the minimum processing power required by 
the baseband signal processors of the OFDM, OFDM-IM, and DM-OFDM transmitters
are respectively given by
\begin{eqnarray} 
\mathcal{I}_{\text{tx-ofdm}} &=& \frac{T_{\text{txofdm}}(64)}{4}=\frac{64\log_2 64}{4} \nonumber\\
&=& 96 \quad \text{inst/$\mu$s,}\\
\mathcal{I}_{\text{tx-im}} &=& \mathcal{I}_{\text{tx-dm}}=\frac{T_{\text{txim}}(64)}{4}=\frac{64\log_2 64+64^2}{4}\nonumber\\ 
&=& 1120 \quad \text{inst/$\mu$s.}\label{eq:Itxdm}
\end{eqnarray}
As expected, the IM transmitters require more powerful baseband signal 
processors with 1120 inst/$\mu$s to meet a maximum processing runtime of 4~$\mu$s. 
This stems from the fact that they rely on the same IxS procedure at the transmitter.
As we will see later, the difference between the complexities of OFDM-IM and DM-OFDM 
lies in the receiver. 

Following similar approach, we refer to~\ref{eq:trxofdm},
\ref{eq:trxim}, and \ref{eq:trxdm} to obtain the required processing power
for the OFDM, OFDM-IM, and DM-OFDM receivers, namely,
\begin{eqnarray} 
\mathcal{I}_{\text{rx-ofdm}} &=& \frac{T_{\text{rxofdm}}(64)}{4} =\frac{64\log_2 64+64}{4}\nonumber\\
&=& 112 \quad \text{inst/$\mu$s,}\\
\mathcal{I}_{\text{rx-im}} &=& \frac{T_{\text{rxim}}(64)}{4}=\frac{64\log_2 64+2\cdot 64 + 64^2}{4}  \nonumber\\
    &=& 1152 \quad \text{inst/$\mu$s,}\\
\mathcal{I}_{\text{rx-dm}} &=& \frac{T_{\text{rxdm}}(64)}{4} =\frac{64\log_2 64+4\cdot 64 + 64^2}{4} \nonumber \\
&=& 1184 \quad \text{inst/$\mu$s.}\label{eq:Irxdm}
\end{eqnarray}
Based on these results, we will assess SC bit rates and SEs considering
the processing power of 1120 inst/$\mu$s and 1184 inst/$\mu$s for all transmitters and
receivers, respectively.
}

\begin{table}[]
\scriptsize
{\color{black}
\caption{\label{tb:results} SC performance indicators of OFDM, OFDM-IM, and DM-OFDM under IEEE 802.11a
parameters ($N$=64, $T_{\text{sym}}=4$~$\mu$s, $W$=20 MHz).}
\begin{tabular}{|l|l|c|c|c|}
\hline
                                                                       & \multicolumn{1}{c|}{\begin{tabular}[c]{@{}c@{}}Performance \\ Metric\end{tabular}}                                                            & \multicolumn{1}{l|}{OFDM} & \multicolumn{1}{l|}{OFDM-IM} & \multicolumn{1}{l|}{DM-OFDM} \\ \hline
\multirow{3}{*}{Tx.}                                                   & \begin{tabular}[c]{@{}l@{}}Computational \\ Complexity\\ $T_{\text{tx}}(64)$ (inst)\end{tabular}                                              & 384                       & 4480                         & 4480                         \\ \cline{2-5} 
                                                                       & \begin{tabular}[c]{@{}l@{}}Minimum Required\\ Processing Power $\mathcal{I}_{\text{tx}}$ (inst/$\mu$s)\end{tabular}                           & 96                        & 1120                         & 1120                         \\ \cline{2-5} 
                                                                       & \begin{tabular}[c]{@{}l@{}}Signal processing runtime \\ under equitable computational \\ resources $T_{\text{comp-tx}}$ ($\mu$s)\end{tabular} & 0.34                      & 4                            & 4                            \\ \hline
\multirow{3}{*}{Rx.}                                                   & \begin{tabular}[c]{@{}l@{}}Computational \\ Complexity\\ $T_{\text{rx}}(64)$ (inst)\end{tabular}                                              & 448                       & 4608                         & 4736                         \\ \cline{2-5} 
                                                                       & \begin{tabular}[c]{@{}l@{}}Minimum Required\\ Processing Power $\mathcal{I}_{\text{rx}}$ (inst/$\mu$s)\end{tabular}                           & 112                       & 1152                         & 1184                         \\ \cline{2-5} 
                                                                       & \begin{tabular}[c]{@{}l@{}}Signal processing runtime \\ under equitable computational \\ resources $T_{\text{comp-rx}}$ ($\mu$s)\end{tabular} & 0.37                      & 3.89                         & 4                            \\ \hline
\multirow{2}{*}{\begin{tabular}[c]{@{}l@{}}L\\ I\\ N\\ K\end{tabular}} & \begin{tabular}[c]{@{}l@{}}Complexity-\\ Constrained\\ Bit rate $\mathsf{SC}_{\text{R}}$\\ (b/$\mu$s)\end{tabular}                            & 10.19                     & 5.71                         & 7.6                          \\ \cline{2-5} 
                                                                       & \begin{tabular}[c]{@{}l@{}}Complexity-\\ Constrained\\ SE $\mathsf{SC}_{\text{SE}}$\\ (b/s/Hz)\end{tabular}                                   & 0.51                      & 0.28                         & 0.38                         \\ \hline
\end{tabular}
}
\end{table}

\subsubsection{Baseband Signal Runtime under Equitable Computational Resources}
To ensure a fairer bit rate analysis, we consider the computation time of each
waveform assuming equitable computational resources of the most complex signal.
According to our prior analysis, both OFDM-IM and DM-OFDM are the most complex transmitters, 
requiring $1120$ inst/$\mu$s each \ref{eq:Itxdm}. If empowered by this superior 
amount of computational resources of DM-OFDM {\color{black} instead of the minimum required by
OFDM (thereby, the denominator of~\ref{eq:tcomptxofdm} modifies accordingly), 
the baseband processor of the OFDM transmitter achieves a runtime performance of
\begin{eqnarray}
T_{\text{comp-txofdm}}=\frac{64\log_2 64}{1120} \approx 0.34\quad \text{$\mu$s},
\end{eqnarray}
whereas the IM transmitters reach a runtime of $4$~$\mu$s, as previously
dimensioned.
}

 Applying similar methodology to the receivers, each waveform must be empowered by a baseband
processor capable of performing $1184$ inst/$\mu$s, which corresponds to the
processing power demanded by DM-OFDM, the most complex receiver. Under such superior
processing power, the computation runtime of the OFDM \ref{eq:tcomprxofdm}, 
OFDM-IM \ref{eq:tcomprxim}, and DM-OFDM \ref{eq:tcomprxdm} at the receivers 
{\color{black} are respectively given by
\begin{eqnarray}
T_{\text{comp-rxofdm}}&=& \frac{64\log_2 64+64}{1184} \approx 0.37 \quad \text{$\mu$s}, \\
T_{\text{comp-rxim}}&=&\frac{64\log_2 64+2\cdot 64+64^2}{1184} \approx 3.89 \text{$\mu$s},\\ 
T_{\text{comp-rxdm}}&=&\frac{64\log_2+4\cdot 64+64^2}{1184} = 4 \quad \text{$\mu$s}.
\end{eqnarray}
As one can see, OFDM has a runtime efficiency one order of magnitude better
than its IM counteparts if provided the same computational resources.

\subsubsection{Discussion}
Assuming each waveform is allowed a maximum signal processing runtime equal 
to the symbol period ($4~\mu$s), the results presented in the previous section 
indicate that the computational resources of IM waveforms are overdimensioned 
if considered for the processing needs of OFDM. Nevertheless, an OFDM chipset designer 
should take these resources into account for fair comparison with the considered IM waveforms.

With these excess computational resources, the OFDM chipset could accommodate more robust configurations or advanced signal processing algorithms, ultimately improving the system's overall bit rate. For example, a larger budget for the baseband processor chipset could enable the implementation of a more efficient bit encoder to enhance error recovery at the receiver. In the case of convolutional encoders, this could involve increasing the constraint length. Similarly, the receiver could adopt more complex algorithms, such as improved channel estimators, equalizers, and decoders, to demodulate more bits efficiently.
}

 \subsubsection{Fairer Bit Rate and SE Analyses}
\begin{figure}[t]
\centering
  \includegraphics[scale=0.65]{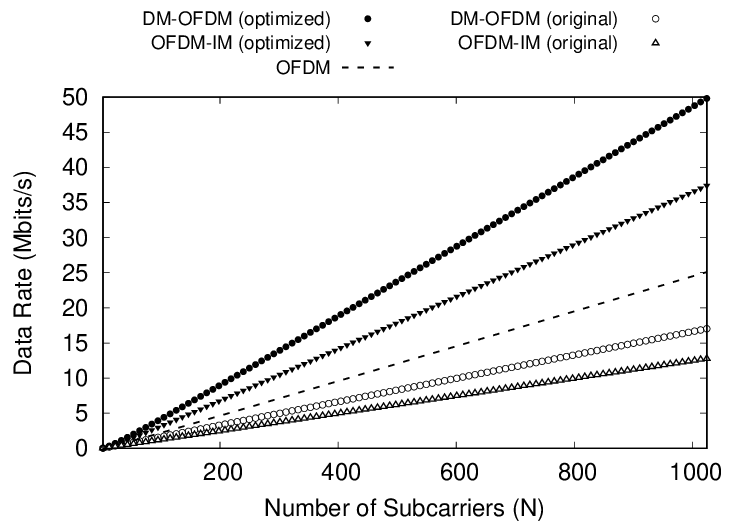}
  \caption{
Computational complexity-constrained bit rate comparison of
original DM-OFDM and OFDM-IM (blank circle and triangle points,
respectively) and OFDM (dashed line).
By providing all baseband processors with the same computational resources of DM-OFDM 
(the most complex waveform), the original IM waveforms can be outperformed by OFDM 
if the optimal complexity-SE balance of DM-OFDM and OFDM-IM (black circle and triangle points,
respectively) is not considered.
}
\label{fig:sc}
\end{figure}

\begin{figure}[t]
\centering
  \includegraphics[scale=0.65]{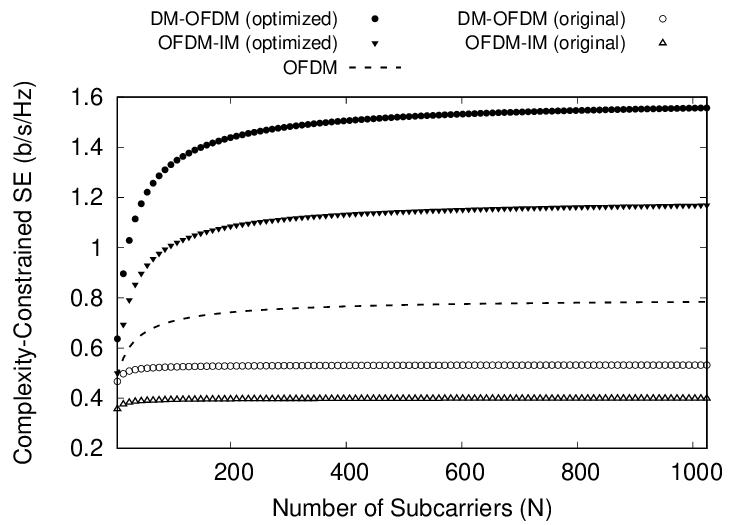}
  \caption{Equivalent time complexity-constrained SE performance of bit rates shown in Fig.~\ref{fig:sc}.}
\label{fig:scse}
\end{figure}

{\color{black}
Given the excessive computational resources of DM-OFDM, an
OFDM designer may consider improving the runtime performance
rather than accommodating more robust algorithms, as previously
discussed. This would reflect in superior SC bit rates and SE.
Indeed, recalling that only 48 (out of 64 subcarriers) convey
data bits in the IEEE 802.11a standard, under the improved processing 
runtime discussed in the prior section, the SC bit rates of OFDM \ref{eq:scofdm}, 
OFDM-IM \ref{eq:scim}, and DM-OFDM \ref{eq:scdm} are
\begin{eqnarray} 
\mathsf{SC}_{\text{R-ofdm}}(48)&=&\frac{48}{0.34+4+0.37} \approx 10.19 \quad \text{b/$\mu$s},\\
\mathsf{SC}_{\text{R-im}}(48)&=&\frac{48/2+\lfloor \log_2 {48\choose 24}\rfloor}{4+4+3.89}\approx 5.71 
\quad \text{b/$\mu$s},\\
\mathsf{SC}_{\text{R-dm}}(48)&=&\frac{48+\lfloor\log_2 {48\choose 24}\rfloor)}{4+4+4} \approx 7.6 \quad
\text{b/$\mu$s,}
\end{eqnarray} 
respectively. This yields respective complexity-constrained SEs of
\begin{eqnarray} 
\mathsf{SC}_{\text{SE-ofdm}}(48)&=&\frac{10.19\text{ b/$\mu$s}}{20 \text{ MHz}} \approx 0.51 \quad \text{b/s/Hz},\\
\mathsf{SC}_{\text{SE-im}}(48)&=&\frac{5.71\text{ b/$\mu$s}}{20 \text{ MHz}}\approx 0.28
\quad \text{b/s/Hz},\\
\mathsf{SC}_{\text{SE-dm}}(48)&=&\frac{7.6 \text{ b/$\mu$s}}{20 \text{ MHz}} \approx 0.38 \quad
\text{b/s/Hz}.
\end{eqnarray} 
These results, along with their previously discussed components, are summarized in Table~\ref{tb:results}. 

Note that the increase in complexity for achieving a higher bit rate 
benefits DM-OFDM more than OFDM-IM. Nonetheless, the low complexity of 
OFDM ensures robust overall performance, provided computational complexity
 is not overlooked and computational resources are equitably considered for
all waveforms. To revert these results, IM waveforms should definitely
adopt more efficient procedures. For example, by considering the optimized 
IxS mapper of~\cite{queiroz-wcl-19},~\cite{queiroz-access-2020} rather than
the originally proposed for the waveforms,
both OFDM-IM and DM-OFDM can achieve better SC bit rates and SE, as shown
in plots of Figs.~\ref{fig:sc} and~\ref{fig:scse} for varying $N$, respectively.
}


\section{Conclusion}\label{sec:conclusion}
In this work, we presented a tutorial on the computational 
complexity-constrained SE analysis of wireless signal waveforms. 
Our tutorial can be applied to any waveform design, regardless of the baseband processor technology, as long as the signal's SE, computational complexity, and symbol period are known.
Through step-by-step examples, we demonstrated how the computational complexity overhead of the physical layer can effectively impact the SE delivered to the upper layers.
This makes a strong case to reconsider the assumption of negligible signal processing
 runtime in the SE analysis, particularly when comparing waveforms that 
present different computational resource constraints to ensure a baseband processor runtime 
below the symbol period. 
We demonstrated this point by a case study in which waveforms
considered less efficient in terms of SE can outperform their
more computationally expensive counterparts when provided
with equivalent high-performance computational resources.

Future works can study the complexity-SE trade-off in the
context of other waveforms such as NOMA and multiple-mode index
modulation. Besides, the SC efficiency analysis can be enhanced
to account for the power consumption of the signal computational
complexity.

\section{Acknowledgements}
Authors would like to thank the Science and Technology Development Fund, Macau SAR. (File no. 0044/2022/A1)
and Agenda Mobilizadora Sines Nexus (ref. No. 7113), supported by the Recovery and Resilience Plan (PRR) and by the European Funds Next Generation EU.

\bibliographystyle{IEEEtran}
\bibliography{IEEEabrv,refs}

\begin{IEEEbiography}[{\includegraphics[width=1in,height=1.25in,clip,keepaspectratio]{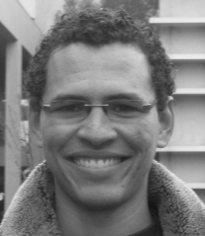}}]{Saulo Queiroz} 
is a professor at the Department of Computer Science of the Federal University of Technology (UTFPR) in Brazil. He completed his Ph.D. with distinction and honor at the University of Coimbra (Portugal). During his academic graduation, he has contributed to open source projects in the field of networking, having participated in initiatives such as as Google Summer of Code. Over the last decade, he has lectured disciplines on computer science such as design and analysis of algorithms, data structures and communication signal processing. His current research interest comprises networking and signal processing for wireless communications.
 \end{IEEEbiography}
\begin{IEEEbiography}[{\includegraphics[width=1in,height=1.25in,clip,keepaspectratio]{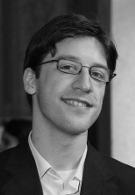}}]{Jo\~ao P. Vilela} 
is an assistant professor at the Department of Computer Science of the University of Porto, Portugal, and a senior researcher at CISUC and INESC TEC. He was a professor at the University of Coimbra after receiving his Ph.D. in Computer Science from the University of Porto in 2011, and a visiting researcher at Georgia Tech and MIT, USA. In recent years, Dr. Vilela has been coordinator and team member of several national, bilateral, and European-funded projects in security and  privacy. His main research interests are in security and privacy of computer and communication systems, with applications such as wireless networks, Internet of Things and mobile devices. Specific research topics include wireless physical-layer security, security of next-generation networks, privacy-preserving data mining, location privacy and automated privacy protection.
\end{IEEEbiography}

\begin{IEEEbiography}[{\includegraphics[width=1in,height=1.25in,clip,keepaspectratio]{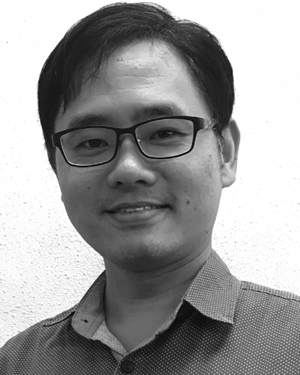}}]{Benjamin K. Ng}
received the B.A.Sc., M.A.Sc., and Ph.D. degrees in engineering science and electrical engineering from the University of Toronto, in 1996, 1998, and 2002, respectively. From 2005 to 2009, he was with Radiospire Networks Inc., Boston, MA, USA, where he was a Senior Communications Engineer focusing on the UWB and millimeter wave technologies. He joined Macao Polytechnic University, Macau, China, in 2010, where he is currently an Associate Professor with the Faculty of Applied Sciences. His research interests include wireless communications and signal processing, with an emphasis on MIMO, NOMA, and machine learning technologies.
\end{IEEEbiography}

\begin{IEEEbiography}[{\includegraphics[width=1in,height=1.25in,clip,keepaspectratio]{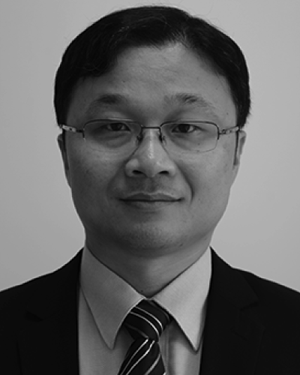}}]{Chan-Tong Lam}
received the B.Sc. (Eng.) and M.Sc. (Eng.) degrees from Queen’s University, Kingston, ON, Canada, in 1998 and 2000, respectively, and the Ph.D. degree from Carleton University, Ottawa, ON, Canada, in 2007. He is currently an Associate Professor with the Faculty of Applied Sciences, Macao Polytechnic University, Macau, China. From 2004 to 2007, he participated with the European Wireless World Initiative New Radio (WINNER) Project. His research interests include mobile wireless communications, machine learning in communications, and computer vision in smart city.
\end{IEEEbiography}

\begin{IEEEbiography}[{\includegraphics[width=1in,height=1.25in,clip,keepaspectratio]{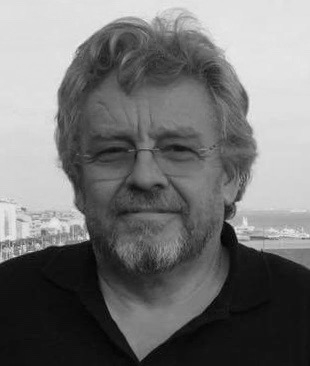}}]{Edmundo Monteiro}
is currently a Full Professor with the University of Coimbra, Portugal.
He has more than 30 years of research experience in the field of computer communications,
wireless networks, quality of service and experience, network and service management, and
computer and network security. He participated in many Portuguese, European, and international
research projects and initiatives. His publication
list includes over 200 publications in journals,
books, and international refereed conferences. He has co-authored nine
international patents. He is a member of the Editorial Board of Wireless
Networks (Springer) journal and is involved in the organization of many
national and international conferences and workshops. He is also a Senior
Member of the IEEE Communications Society and the ACM Special Interest
Group on Communications. He is also a Portuguese Representative in IFIP
TC6 (Communication Systems).
\end{IEEEbiography}

\end{document}